\documentclass{emulateapj}

\begin{document}

\shorttitle{Mass Loss in NGC~6791}
\shortauthors{Kalirai et al.}

\title{Stellar Evolution in NGC~6791: Mass Loss on the Red Giant Branch \\ and 
the Formation of Low Mass White Dwarfs\altaffilmark{1,2}}

\author{
Jasonjot S. Kalirai\altaffilmark{3,4}, 
P. Bergeron\altaffilmark{5},
Brad M.~S. Hansen\altaffilmark{6},
Daniel D. Kelson\altaffilmark{7}, \\
David B. Reitzel\altaffilmark{6},
R. Michael Rich\altaffilmark{6}, and
Harvey B. Richer\altaffilmark{8}
}
\altaffiltext{1}{Data presented herein were obtained at the W.\ M.\ Keck
Observatory, which is operated as a scientific partnership among the
California Institute of Technology, the University of California, and the
National Aeronautics and Space Administration.  The Observatory was made
possible by the generous financial support of the W.\ M.\ Keck Foundation.}
\altaffiltext{2}{Based on observations obtained at the Canada-France-Hawaii 
Telescope (CFHT) which is operated by the National Research Council of Canada, 
the Institut National des Sciences de l'Univers of the Centre National de 
la Recherche Scientifique of France, and the University of Hawaii.}
\altaffiltext{3}{University of California Observatories/Lick Observatory, 
University of California at Santa Cruz, Santa Cruz CA, 95060; jkalirai@ucolick.org}
\altaffiltext{4}{Hubble Fellow}
\altaffiltext{5}{D\'epartement de Physique, Universit\'e de Montr\'eal, C.P.~6128, 
Succ.~Centre-Ville, Montr\'eal, Qu\'ebec, Canada, H3C 3J7; bergeron@astro.umontreal.ca}
\altaffiltext{6}{Department of Physics and Astronomy, Box 951547, Knudsen Hall, 
University of California at Los Angeles, Los Angeles CA, 90095; 
hansen/rmr/reitzel@astro.ucla.edu}
\altaffiltext{7}{Carnegie Observatories, Carnegie Institution of Washington, 813 Santa 
Barbara Street, Pasadena CA, 91101; kelson@ociw.edu}
\altaffiltext{8}{Department of Physics and Astronomy, University of British Columbia, 
Vancouver, British Columbia, Canada, V6T 1Z1; richer@astro.ubc.ca}


\begin{abstract}
We present the first detailed study of the properties (temperatures, gravities, 
and masses) of the NGC~6791 white dwarf population.  This unique stellar system 
is both one of the oldest (8~Gyr) and most metal-rich ([Fe/H] $\sim$ $+$0.4) open 
clusters in our Galaxy, and has a color-magnitude diagram (CMD) that exhibits 
both a red giant clump and a much hotter extreme horizontal branch.  Fitting the 
Balmer lines of the white dwarfs in the cluster, using Keck/LRIS spectra, suggests 
that most of these stars are undermassive, 
$\langle$$M$$\rangle$ = 0.43 $\pm$ 0.06 $M_\odot$, and therefore could not 
have formed from canonical stellar evolution involving the helium flash 
at the tip of the red giant branch.  We show that at least 40\% of NGC~6791's 
evolved stars must have lost enough mass on the red giant branch to 
avoid the flash, and therefore did not convert helium into carbon-oxygen 
in their core.  Such increased mass loss in the evolution of the progenitors 
of these stars is consistent with the presence of the extreme horizontal 
branch in the CMD.  This unique stellar evolutionary channel also naturally 
explains the recent finding of a very young age (2.4~Gyr) for NGC~6791 from 
white dwarf cooling theory; helium core white dwarfs in this cluster will 
cool $\sim$3 times slower than carbon-oxygen core stars and therefore the 
corrected white dwarf cooling age is in fact $\gtrsim$7~Gyr, consistent with 
the well measured main-sequence turnoff age.   These results provide 
direct empirical evidence that mass loss is much more efficient in high 
metallicity environments and therefore may be critical in interpreting the 
ultraviolet upturn in elliptical galaxies.

\end{abstract}



\section{Introduction} \label{introduction}

The advent of wide-field imaging cameras on 4-meter class telescopes 
in the last decade has led to a number of large homogeneous surveys of 
the Galactic open star cluster population.  These studies not only shed 
light on cluster properties (e.g., distances, stellar content, ages, and 
metallicities) and their formation mechanisms, they also directly help in 
our understanding of the evolution and chemical structure of the Galactic 
disk.  For example, the WIYN Open Cluster Survey \citep{mathieu00} and the CFHT 
Open Cluster Survey \citep{kalirai01a} have now targeted several dozens of these 
systems in our Galaxy (e.g., NGC~6819 -- Kalirai et~al.\ 2001b; NGC~2099 -- Kalirai 
et~al.\ 2001c; NGC~2168 and NGC~2323 -- von Hippel et~al.\ 2002 and Kalirai 
et~al.\ 2003).   

NGC~6791 is a relatively nearby star cluster ($d \sim$ 4 kpc -- Chaboyer, 
Green, \& Liebert 1999) located at ($l$, $b$) = (69.96$^\circ$, 
10.90$^\circ$).  Very early studies of the system established it as one 
of the most populous open star clusters, with a mass of several thousand 
Solar masses (e.g., Kinman 1965).  These first studies also concluded that 
NGC~6791's stellar content is both very old and has a high metal abundance 
(e.g., Spinrad \& Taylor 1971).  More recent studies have confirmed these 
earlier results with greater precision; current best estimates indicate that 
the age of NGC~6791 is $\gtrsim$8~Gyr, the [$\alpha$/Fe] is Solar 
\citep{origlia06}, and the metallicity is [Fe/H] = $+$0.3 -- $+$0.5 
\citep{kaluzny90,demarque92,montgomery94,peterson98,chaboyer99,
stetson03,carney05,gratton06,origlia06}.  The cluster therefore ranks as 
both one of the oldest open clusters and one of the most metal-rich in 
our Galaxy \citep{friel93}.  Given this unique combination, 
NGC~6791 currently serves as {\it the} high metallicity anchor when measuring 
star formation histories from CMDs of nearby 
galaxies. 

The CMD of NGC~6791 exhibits some peculiar features (e.g., Stetson et~al.\ 2003).  
The cluster contains a large blue straggler population, and both a red giant 
clump and an extremely blue horizontal branch.  Given 
the high metallicity, this is a strong example of the second parameter 
effect.  The extreme horizontal branch has very likely formed as a result of 
increased mass loss in post main-sequence evolutionary phases, possibly due 
to the high metallicity of the cluster \citep{faulkner72,sweigart87,castellani93}.  Although the 
presence of these stars in the field has been suggested to possibly arise 
from binary evolution (e.g., Allard et~al.\ 1994; Maxted et~al.\ 2001; 
Han et~al.\ 2003), this does not appear to be the case in star clusters 
(e.g., Moni~Bidin et~al.\ 2006a), especially a system like NGC~6791 (see 
discussion in \S\,\ref{ehbstars}).  The cluster orbit is highly eccentric, 
which combined with its chemical content and position, has led to suggestions 
that it may even represent the nucleus of a tidally disrupted galaxy 
\citep{carraro06}.  The unique properties of NGC~6791 certainly hold 
promising information on its origins and past dynamical and stellar 
evolutionary history. 

Recently, \cite{king05} produced the deepest CMD for NGC~6791 to 
date.  Using the Hubble Space Telescope ({\it HST}) Advanced Camera for Surveys, 
they observed the cluster for 4 orbits, reaching a limiting 
magnitude of $F606W$ = 28.  The resulting CMD shows a tightly constrained main-sequence 
to the limit of the data and, for the first time, has uncovered a large population of 
hundreds of white dwarfs in the cluster \citep{bedin05}.  These stellar remnants 
are cooling with age, becoming fainter as time passes, and therefore serve as 
{\it clocks} from which the cluster can be dated (see e.g., Hansen et~al.\ 2004 
for a detailed discussion).  This technique of determining ages of star clusters 
from white dwarf cooling theory successfully reproduces independently measured 
main-sequence turnoff ages in the six other open clusters, and two globular clusters, 
that have been tested to date \citep{vonHippel05,hansen04,hansen07}.  However, 
\cite{bedin05} conclude that the white dwarf cooling age of NGC~6791 is in fact 
2.4~Gyr, a factor of three less than the well measured main-sequence turnoff age.  
Such a discrepancy clearly adds to the list of peculiarities of this cluster.


\begin{figure}
\epsscale{1.1} \plotone{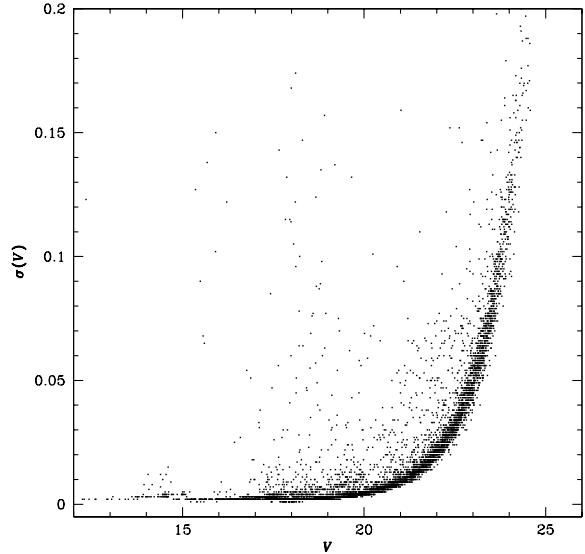} \figcaption{The DAOPHOT photometric error,  
as a function of $V$ magnitude indicates the photometry is accurate to 
$V \gtrsim$ 24 (where the error is $\sigma_V$ = 0.12 magnitudes).
\label{fig:errfig}}
\end{figure}


In this paper we present evidence that the white dwarf population of NGC~6791 is 
unlike that in other clusters.  The formation of most of these stars has resulted 
from a unique evolutionary channel involving significant mass loss on the red giant 
branch, leading to a final mass below the critical mass needed to ignite helium 
in the core of the star \citep{hansen05}.  Hence, the progenitors of these white 
dwarfs avoided the helium flash and therefore the cores of the white dwarfs are 
composed of helium and not carbon-oxygen.  As a result, the masses of the white 
dwarfs are well below the expected 0.5 -- 0.6 $M_\odot$ value that the canonical 
channel produces for these initial masses.  Invoking helium core white dwarf 
models \citep{hansen05} in the fit of the white dwarf cooling sequence from 
\cite{bedin05} yields a consistent age for the cluster as measured from the 
turnoff.  

In the next section, 
we discuss our imaging data set for NGC~6791.  We present a new CMD of the 
cluster in \S\,\ref{CMD}, discuss its various features, and estimate an age for 
the cluster from the new data.  Next, we summarize the findings of \cite{bedin05} 
and consider possible explanations in \S\,\ref{bedinage}.  The first spectroscopic 
observations of NGC~6791's white dwarf population are presented in 
\S\,\ref{spectroscopicdata} and \S\,\ref{WDspectra} and temperatures, gravities, 
and masses for these stars are derived in \S\,\ref{WDMasses}.  The results and their 
implications are discussed in \S\,\ref{discussion} and the study is summarized 
in \S\,\ref{conclusion}.


\section{Imaging Observations} \label{observations}

We imaged NGC~6791 with the CFH12K mosaic CCD on the Canada-France-Hawaii 
Telescope (CFHT) in March and April of 2001.  This camera contains 12 CCDs, 
each with 2048 $\times$ 4096 pixels, where each pixel subtends 0$\farcs$206.  
The detector projects to an area of 42$'$ $\times$ 28$'$ on the sky, much larger 
than the size of the cluster.  The observations were taken in the $B$ and $V$ 
bands with the center of the cluster placed on one of the CCDs 
(away from the center of the mosaic where stars would be lost due 
to chip gaps).  Seven exposures were taken in each filter (each one 850 seconds 
in $V$ and 1150 seconds in $B$) to achieve a photometric depth fainter than 
$B, V \sim$ 24 over a magnitude fainter than the brightest expected 
cluster white dwarfs.  Shallower exposures were also obtained to 
obtain photometry of the brighter stars that were saturated on the deeper exposures.  
Most observations were obtained in sub-arcsecond seeing and all were 
taken under photometric skies.  Table~1 presents a complete observational 
log of the imaging data.


\begin{table}
\begin{center}
\caption{}
\begin{tabular}{lcccr}
\hline
\hline
\multicolumn{1}{c}{Filter} & \multicolumn{1}{c}{Exp. Time (s)} & 
\multicolumn{1}{c}{No. Images}  & \multicolumn{1}{c}{Seeing ($''$)} & 
\multicolumn{1}{c}{Airmass} \\ 
\hline
$V$ & 850  & 7 & 0.63 -- 0.98 & $<$1.25 \\
$V$ & 300  & 1 & 0.93       & 1.25 \\
$V$ & 90   & 1 & 1.03       & 1.26 \\ 
$V$ & 20   & 1 & 0.86       & 1.09 \\
$V$ & 10   & 1 & 0.92       & 1.26 \\
$V$ & 5    & 1 & 1.03       & 1.28 \\
$V$ & 1    & 1 & 1.02       & 1.28 \\
$B$ & 1150 & 7 & 0.87 -- 1.30 & $<$1.21 \\
$B$ & 400  & 1 & 0.86       & 1.14 \\
$B$ & 120  & 1 & 0.98       & 1.15 \\
$B$ & 30   & 1 & 1.07       & 1.09 \\
$B$ & 10   & 1 & 0.75       & 1.12 \\
$B$ & 5    & 1 & 0.73       & 1.12 \\
$B$ & 1    & 1 & 0.79       & 1.12 \\
\hline
\end{tabular}
\label{table1}
\end{center}
\end{table}


The data were processed (flat-field, bias and dark corrected) and
montaged using the FITS Large Images Processing 
Software\footnote{\url{http://www.cfht.hawaii.edu/$\sim$jcc/Flips/flips.html}} 
(FLIPS) as described in Kalirai et~al.\ (2001a). The photometry of 
all sources was performed using a variable point-spread function 
in DAOPHOT (Stetson 1994).  The photometry was calibrated using 
Landolt standard star fields as discussed in \S\S 5.1 and 5.2 of 
\cite{kalirai01a}.  The mean errors in the photometry are 
$\sigma_V$ = 0.02 mag at $V$ = 22, $\sigma_V$ = 0.05 mag at 
$V$ = 23, and $\sigma_V$ = 0.12 mag at $V$ = 24.  A statistical 
error plot for several thousand stars in the vicinity of the cluster 
is shown in Figure~\ref{fig:errfig}.

Figure~\ref{fig:starcountmap} shows a starcount map constructed from our 
CFHT imaging observations.  We have included all objects within a generous 
envelope of the cluster main sequence on the CMD (see \S\,\ref{CMD}).  With 
this mild cut, NGC~6791 stands out very strongly against the 
foreground/background Galactic disk stars.  The rectangular region marks 
the Keck LRIS field of view over which we obtained spectroscopy of 
white dwarf candidates (see \S\,\ref{spectroscopicdata}).


\section{The Color-Magnitude Diagram of NGC~6791} \label{CMD}

\begin{figure}
\epsscale{1.1} \plotone{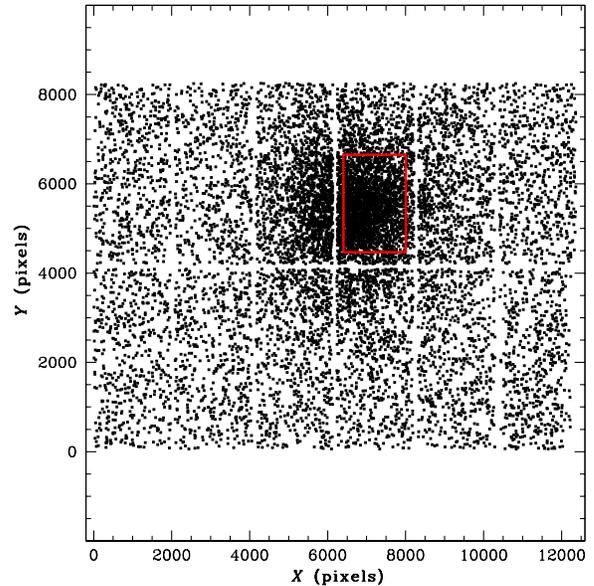} 
\figcaption{A wide-field starcount map of NGC~6791 constructed from the 
CFHT imaging.  A mild cut has been used to isolate stars within an 
envelope of the cluster main sequence.  The region in which spectroscopic 
targets were selected for Keck/LRIS observations is indicated with a 
rectangle (see \S\,\ref{spectroscopicdata}).
\label{fig:starcountmap}}
\end{figure}

\begin{figure*}
\begin{center}
\leavevmode 
\includegraphics[height=15cm,width=13cm,angle=270]{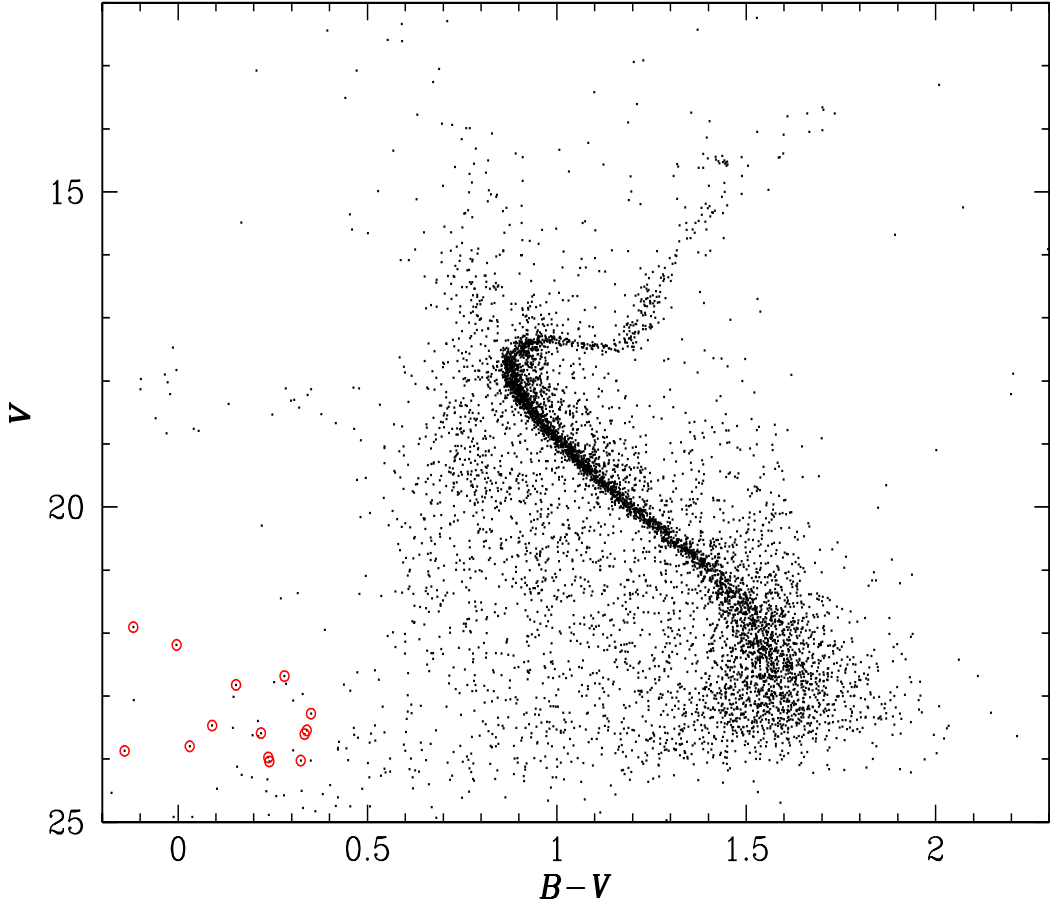}
\end{center}
\caption{The CMD of NGC~6791 from our CFHT CFH12K imaging data.  
A very tight cluster main-sequence, and several post main-sequence 
evolutionary phases can be clearly seen.  Roughly a dozen bright, 
extremely blue horizontal branch stars are also evident at 
$B{\rm-}V \sim$ 0, $V \sim$ 17.  The faint, blue region of the CMD 
shows several potential white dwarf candidates.  The 14 objects that 
were targeted with Keck/LRIS are highlighted with larger open circles 
(see \S\,\ref{spectroscopicdata}).
\label{fig:cmd}}
\end{figure*}

The CMD for NGC~6791 is presented in Figure~\ref{fig:cmd} for all 
stars that fall within an area slightly larger than the Keck LRIS 
spectroscopic mask shown in Figure~\ref{fig:starcountmap}.  The CMD 
clearly shows all of the major phases of stellar evolution: 
the main-sequence, turnoff, subgiant branch, red giant branch, and 
red giant clump.  A significant population of potential blue straggler 
stars is also seen above the cluster turnoff.

The red giant clump of NGC 6791 represents a phase of core helium 
burning following the helium flash at the tip of the cluster's 
red giant branch.  The result of this burning is a star with a 
carbon-oxygen core.  As has been noted in earlier studies (e.g., 
Kaluzny \& Udalski 1992; Liebert, Saffer, \& Green 1994; Kaluzny \& 
Rucinski 1995; Green, Liebert, \& Peterson 1996), the NGC~6791 CMD 
also shows about a dozen extreme horizontal branch stars 
(at $B{\rm-}V \sim$ 0, $V \sim$ 17), most of which are likely 
subdwarf B and subdwarf O stars.  Although these much hotter 
stars are also burning helium in their cores, their evolution has 
differed from the red giant clump stars.  These stars likely 
represent the products of increased mass loss on the red giant 
branch \citep{faulkner72} and possibly suffered a delay in the 
ignition of the core helium in the star until a point where the 
star contracted further \citep{lanz04,castellani93}.  In this picture of single star 
evolution, it is believed that the high metallicity of the cluster is 
driving the enhanced mass loss (e.g., D'Cruz et~al. 1996).  
\cite{yong00} also consider whether mass loss on the horizontal branch 
itself could have led, in part, to the morphology of the extreme 
horizontal branch of this cluster.  


In the faint-blue part of the CMD in Figure~\ref{fig:cmd} we see a 
population of white dwarf candidates.  Given the richness of NGC~6791 
and the position of our spectroscopic mask (see 
Figure~\ref{fig:starcountmap}), we statistically expect most of the 
white dwarfs in our sample to be a part of the cluster.  The starcount 
map in Figure~\ref{fig:starcountmap} shows that NGC~6791 is 
centered in the top row CCDs, slightly to the right of the center 
of the camera.  We can directly measure the field white dwarf 
density by examining the faint-blue end of a CMD constructed from 
the outer CCDs in the bottom row.  We take a region with an area 
$\gtrsim$4$\times$ our LRIS field and count a dozen stellar 
objects within the same magnitude and color range that we use 
for selecting white dwarf targets (see \S\,\ref{spectroscopicdata}).  
Scaling by the ratio of areas, the number of field white dwarfs in 
our sample is therefore expected to be $\lesssim$3.


\subsection{Cluster Reddening, Distance, Age, and Metallicity} \label{red.dist.age}

The foreground reddening, distance, age, and metallicity of NGC~6791 have 
been estimated many times in the literature (see references in 
\S\,\ref{introduction}).  Recent values based on {\it HST} filters 
\citep{king05}, $B, V, I$ optical data \citep{chaboyer99,stetson03}, 
and $J, H, K$ near infrared observations \citep{carney05} find E($B-V$) = 
0.09 -- 0.18.  The same studies estimate the distance of NGC~6791 to be 
$d \sim$ 4000~pc (the range in these studies is $d$ = 3600 -- 4200~pc).  
Most determinations of the age of NGC~6791 have resulted 
from fitting theoretical isochrones to the observed cluster main sequence 
and turnoff morphology.  Such determinations are strongly dependent on the 
assumed reddening, distance, and 
metallicity.  Differences in the input physics within various groups theoretical 
models (e.g., helium abundance and treatment of overshooting) also play an 
appreciable role in the age determinations.  Therefore, recent values in the literature 
have ranged from $\sim$8~Gyr (e.g., Chaboyer, Green, \& Liebert 1999) to as high as 
12~Gyr (e.g., Stetson, Bruntt, \& Grundahl 2003).  As we mentioned earlier, the 
cluster has been known to have a high metal abundance for some time.  The first medium 
resolution spectroscopy found [Fe/H] = $+$0.40 $\pm$ 0.10 \citep{peterson98}.  Two very 
recent studies based on high resolution infrared spectroscopy \citep{origlia06} and 
high resolution optical spectroscopy \citep{gratton06} confirm this.  \cite{origlia06} 
find [Fe/H] = $+$0.35 $\pm$ 0.02 and \cite{gratton06} find [Fe/H] = $+$0.47 $\pm$ 0.04.

\begin{figure}
\epsscale{1.1} \plotone{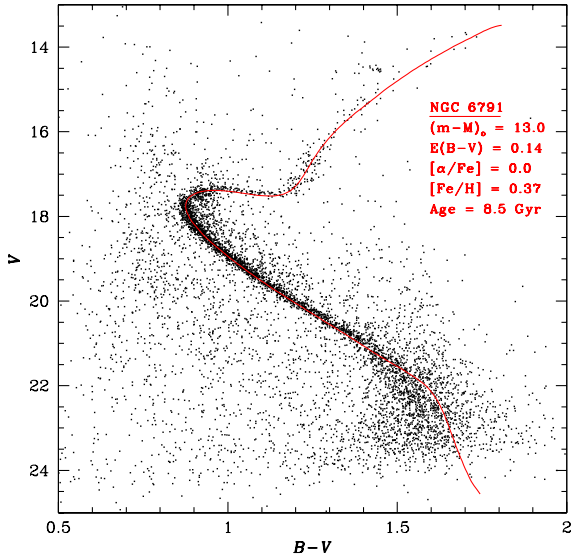} 
\figcaption{An 8.5~Gyr isochrone with [Fe/H] = $+$0.37 \citep{vandenberg05} provides an 
excellent fit to the main-sequence, turnoff, sub-giant branch, and red giant branch 
of NGC~6791.  These data therefore support previous findings that the cluster is 
both very old and metal rich.
\label{fig:iso}}
\end{figure}

Our CMD of NGC~6791 can be used to independently determine the age of the cluster.  
We find that for a choice of E($B-V$) = 0.14 \citep{carney05}, ($m-M$)$_{\rm o}$ = 13.0 
(an average of the four recent studies referenced above), and [Fe/H] = $+$0.37, an isochrone 
with [$\alpha$/Fe] = 0 and age = 8.5 Gyr \citep{vandenberg05} provides an excellent fit to 
the observed CMD.  This is shown in Figure~\ref{fig:iso}.  Adopting a slightly larger 
metallicity (e.g., [Fe/H] = $+$0.47 -- Gratton et~al.\ 2006) requires a younger age by 
$\sim$1~Gyr.  However the fit is significantly worse along the subgiant and red giant 
branches.  Similar variations in the reddening and distance modulus also produce smaller 
age changes.  Therefore, our data supports the literature results that the cluster is 
very old, and metal-rich.  In a future paper, we will provide a full analysis of the 
entire data set in the CFHT mosaic image.  This will include the first determination of 
the cluster's distance, age, reddening, binary fraction, and mass based on 
MonteCarlo simulations of synthetic CMDs.  These comparisons, as shown in 
Kalirai \& Tosi (2004) for several open clusters, allow modeling of several additional 
parameters which dictate the distribution of points in the CMD, such as stochastic 
star formation processes, photometric spread, data incompleteness, and cluster 
luminosity function.






\section{A White Dwarf Cooling Age \\ for NGC~6791 of 2.4 Gyr?} \label{bedinage}

Up until recently, all of the studies that have measured the age of the cluster used 
the same technique, isochrone fitting of the main-sequence turnoff.  Recently, 
\cite{bedin05} have 
imaged NGC~6791 with the {\it HST} Advanced Camera for Surveys down to very 
faint magnitudes (F606W = 28).  Their study was the first to uncover the 
remnant population of evolved stars in the cluster (see their Figure~1).  Since 
these stars have no remaining nuclear energy sources, they cool with time and become 
predictably fainter.  \cite{bedin05} model the observed luminosity function of these 
white dwarfs and provide the first independent age measurement for the cluster.  Given 
the morphology and peak of the observed white dwarf luminosity function, white 
dwarf cooling models from \cite{salaris00} indicate that the cluster is only 2.4 Gyr 
old.  This age is at least a factor of three less than the main-sequence turnoff 
age for the cluster.

\subsection{Possible Explanations} \label{explanations}

\cite{bedin05} consider several explanations for the white dwarf (WD) cooling age discrepancy 
in NGC~6791 but find that none of them are very satisfactory.  These include using 
radically different initial-to-final mass mappings, incorrect distance moduli or 
metallicities, different hydrogen-layer thicknesses for the WDs, and binary evolution.

At least two additional theories have been proposed to explain the above anomalous age 
result that are more promising.  The first suggests that the cooling rate of white 
dwarfs may be retarded in a system such as NGC~6791 given the high metallicity of the 
cluster.  \cite{deloye02} predicted that gravitational settling of $^{22}$Ne would result 
in an increased release of gravitational energy that may not be seen in other less 
metal-rich systems.  In fact, they explicitly say that a cluster such as NGC~6791 is an 
ideal environment to test this effect.  However, the magnitude of the delay is 
predicted to be 0.25 -- 1.6~Gyr (although it does depend on an uncertain diffusion 
coefficient) so it is not clear whether it, {\it or it alone}, can explain the 
observed discrepancy in the turnoff and white dwarf cooling ages of NGC~6791.  
L.~Bildsten (2007, private communication) is in the process of investigating 
this possible explanation further.


\begin{figure*}
\begin{center}
\leavevmode 
\includegraphics[height=17cm,angle=270]{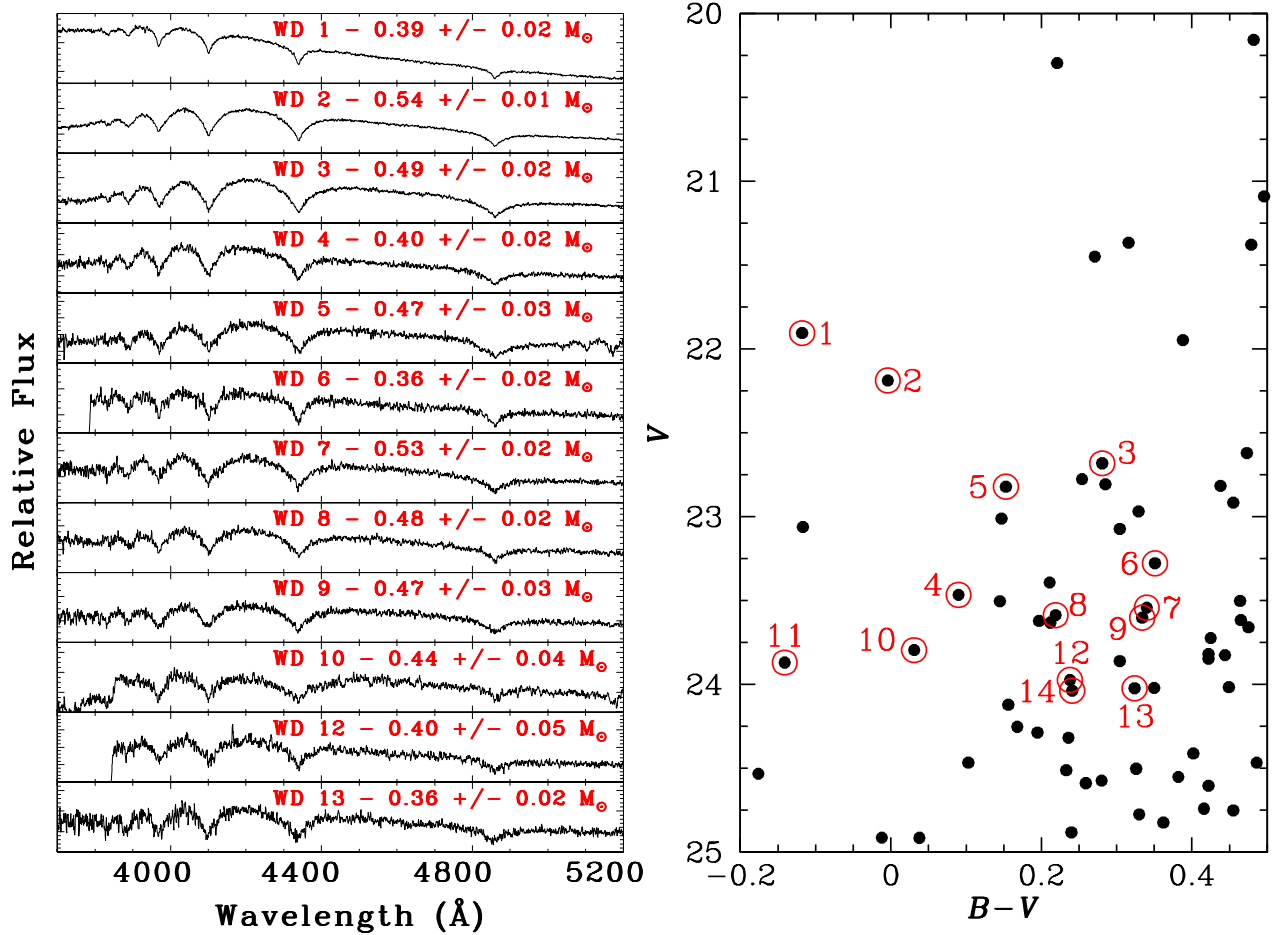}
\end{center}
\caption{{\it Left} - Keck/LRIS spectra confirm that 12 of the 14 
faint-blue targets in our spectroscopic sample are in fact white 
dwarfs. The spectra of these stars show broad hydrogen Balmer 
lines that we fit to model line profiles to derive individual 
stellar masses (indicated within each panel -- see \S\,\ref{WDMasses}).  
The spectra for two of the faintest targets were of poor quality and 
did not permit an accurate classification of the objects. {\it Right} - 
The white dwarf region of the CMD is shown with identifications marking each 
of the spectroscopically targeted stars (larger open circles).  
The identifications are consistent with those in the adjacent 
panel displaying the spectra for these stars.
\label{fig:wdspectra}}
\end{figure*}


The second scenario proposed by \cite{hansen05}, suggests that mass-loss 
on the red giant branch may be the culprit.  Given the higher metallicity 
in NGC~6791, theoretical models of stellar evolution (e.g., Marigo 2001) 
predict that post-main sequence stars in this cluster would lose more 
mass than in less metal-rich systems (see \S\,\ref{theoryRGB}).  If some 
stars can expel enough mass on the red giant branch, they may be peeling 
away towards the white dwarf cooling phase before reaching the helium 
flash.  Therefore, the use of carbon-oxygen core white dwarf models to 
date NGC~6791 will yield an incorrect age measurement.  It is interesting 
to note that a fit to helium core white dwarf models recovers an age 
that is roughly 3$\times$ larger than the \cite{bedin05} result, and 
therefore consistent with the main-sequence turnoff age \citep{hansen05}.  
In the next section, we test this hypothesis.

Although not as extreme a case, it is worth noting that we have seen 
hints for the dependence of mass loss on metallicity in another 
set of clusters.  Both the Hyades \citep{perryman98} and NGC~2099 
(Kalirai et~al.\ 2001c; 2005a) are of similar age, yet their metallicities 
differ by a factor of two ([Fe/H]$_{\rm Hyades}$ = $+$0.17 and 
[Fe/H]$_{\rm NGC~2099}$ = $-$0.1).  An initial-to-final mass relationship 
based on spectroscopically observed white dwarfs in these two clusters 
\citep{claver01,kalirai05a} suggests that stars in NGC~2099, through 
their evolution, have lost less mass than stars in the Hyades.  The 
mean mass of white dwarfs in NGC 2099 is $M$ = 0.80 $\pm$ 0.03 $M_\odot$ 
whereas white dwarfs in the Hyades have $M$ = 0.72 $\pm$ 0.02 $M_\odot$.

\subsubsection{Mass Loss on the Red Giant Branch: \\ Testing the Theory} \label{theorytest}

The presence of two distinct phases of core-helium burning in this cluster 
(the red giant clump and the extreme horizontal branch) hints that mass loss 
is stochastic in this cluster.  For a metallicity of [Fe/H] = $+$0.4, the 
critical mass needed to ignite helium in the core of a star is 0.45 -- 0.47~$M_\odot$ 
(Dominguez et~al.\ 1999; Pietrinferni et~al.\ 2004; VandenBerg, Bergbusch, \& Dowler 
2005; L. Girardi 2006, private communication).  Therefore, a direct prediction of 
\cite{hansen05} is that a large fraction of the white dwarfs along the 
\cite{bedin05} cooling sequence should have masses less than this critical mass.  
Such objects are very rare, both in other star clusters and in the 
field (e.g., from the Palomar Green Survey -- see Liebert, Bergeron, \& Holberg 
2005), and therefore their discovery would almost certainly validate this 
suggestion.


\section{Spectroscopic Observations} \label{spectroscopicdata}

We obtained multi-object spectroscopic observations of the brightest 
white dwarf candidates detected in our CFHT imaging study with 
the Keck~I telescope on 3-4 August 2005.  We designed a single mask and 
targeted 14 objects with the Low Resolution Imaging Spectrometer (LRIS -- Oke 
et~al.\ 1995) over the 5$' \times$ 7$'$ field of view.  These objects were 
selected based on their magnitudes, colors, and location within our much 
larger CFHT field of view.  The spectra were obtained using the 
600/4000 grism which simultaneously covers 2580 ${\rm \AA}$.  The total 
exposure time was 21,600 seconds.  The seeing was variable during the run, 
ranging from 0$\farcs$5 to 1$\farcs$1.  

The spectra were reduced using python routines specifically written for 
LRIS data and are described in detail in \cite{kelson00} and \cite{kelson03}.  
To summarize the key steps, the individual exposures were first bias 
subtracted using the overscan region.  Next, the vertical distortion 
(spatial axis) was measured using cross-correlations along the slit 
edges of the spectropscopic flat-fields, and the boundaries of the 
slitlets were identified using Laplacian edge-detection.  The wavelength 
calibration was performed in an automated way using the Hg, Cd, Zn, 
Ne lamp lines and the zero-points of the dispersion solutions were refined 
using night sky emission lines.  The $rms$ scatter about the dispersion 
solutions was typically $<$0.05 pixels.  The data were corrected for 
pixel-to-pixel variations by dividing  by a normalized spectral 
flat-field.  The spectrum of the night sky was fitted for, and subtracted off, 
using bivariate cubic B-splines fit to the data on both sides of the 
targets.  Finally, one-dimensional spectra were extracted and coadded using 
standard IRAF task and flux calibrated using a spectrophotometric standard 
star (HZ~21).

\section{The Spectra of White Dwarfs in NGC~6791} \label{WDspectra}

In Figure~\ref{fig:wdspectra} (left) we present the optical spectra for 12 of the 
14 faint-blue objects that were targeted with LRIS on Keck~I.  As discussed 
earlier, most of these objects are likely to be cluster members and therefore 
must be white dwarfs.  The spectra confirm this.  All of these objects show 
pressure broadened Balmer lines, from H$\beta$ at 4861 ${\rm \AA}$ to higher order 
lines up to H9 at 3835 ${\rm \AA}$, a clear signature of DA white dwarfs.  The two 
objects not shown (WDs 11 and 14) were among the faintest objects targeted 
and the spectra do not contain enough signal-to-noise to classify the objects.  
The right panel shows the faint-blue region of the cluster CMD with the 12 
objects indicated as large open circles.  The two objects for which the 
spectra are not shown are also indicated. 

Although the Balmer lines are the most prominent features in these white dwarf 
spectra, a closer look reveals other interesting features in two stars.  Towards 
the red end of our spectral coverage for WD~5, we see evidence for additional 
absorption lines.  Similarly, the spectrum of WD~10 shows some contaminating 
lines.  These objects therefore may represent DA+dM binary systems.  Fortunately, 
LRIS is a dual beam spectrograph and therefore we have simultaneous 
observations of these stars extending to beyond 7500 ${\rm \AA}$.  A reduction of 
those data for these targets should reveal any counterparts and certainly 
lead to a better understanding of the nature of these objects. 


\begin{figure*}
\begin{center}
\leavevmode 
\includegraphics[height=17cm,angle=270]{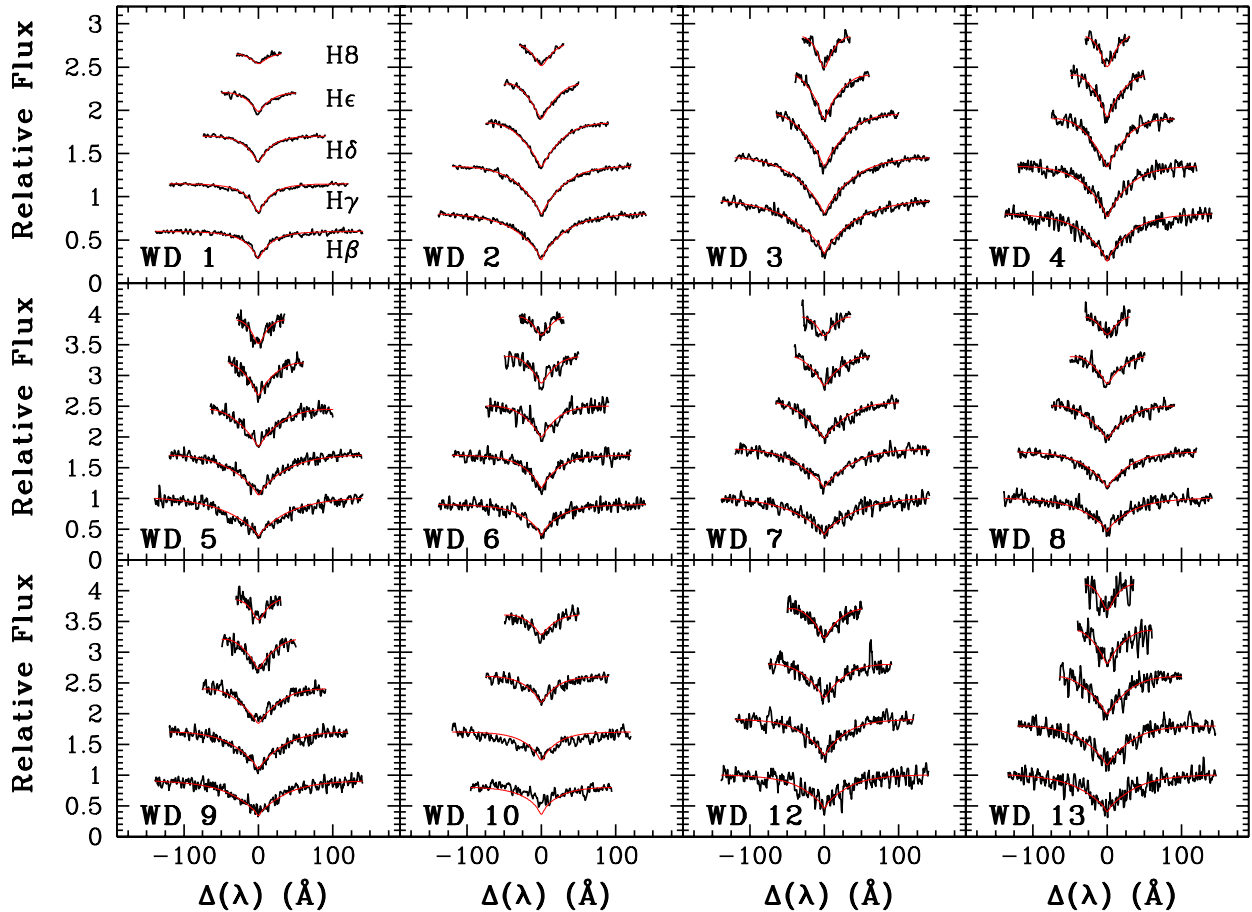}
\end{center}
\caption{Individual hydrogen Balmer lines are shown for 12 white 
dwarfs in NGC~6791 (see Figure~\ref{fig:wdspectra} for 
identifications).  Within each panel, the lines for a given white 
dwarf are H$\beta$ (bottom), H$\gamma$, H$\delta$, H$\epsilon$, 
and H$8$ (top).  Spectroscopic fits, simultaneously to all lines, 
constrain the $T_{\rm eff}$, log~$g$, and mass of each 
white dwarf as discussed in \S\,\ref{WDMasses} (smooth profiles).  
For WD~1 -- WD~4, the H$9$ Balmer line at 3835~${\rm \AA}$ 
was also used in the fits (not shown).  The uncertainties on 
$T_{\rm eff}$ and log~$g$ for WD~10, WD~12, and WD~13 are larger 
than for the other white dwarfs as discussed in the text.  
Table~2 summarizes the results from these fits.
\label{fig:wdmasses}}
\end{figure*}


\subsection{Determining $T_{\rm eff}$, log~$g$, and Masses for 
NGC~6791's White Dwarf Population} \label{WDMasses}

We determine the effective temperatures ($T_{\rm eff}$) and gravities 
(log~$g$) for the twelve white dwarfs shown in Figure~\ref{fig:wdspectra} 
using the techniques described in \cite{bergeron92}.  These parameters 
are calculated for each white dwarf using the nonlinear least-squares 
method of Levenberg-Marquardt \citep{press86}.  For combinations of 
these values, $\chi^{2}$ is minimized using normalized model line 
profiles of {\it all} absorption lines simultaneously.  These fits 
are shown in Figure~\ref{fig:wdmasses}.  For WD~1 -- WD~9, the spectra 
have very well characterized higher order Balmer lines (e.g., at least 
H$8$ and up to H$9$ for four stars -- WD~1, WD~2, WD~3, and WD~4) and the 
model atmosphere fits to all lines are excellent.  For WD~10 and WD~12, 
the spectra are truncated shortward of $\sim$3850~${\rm \AA}$ as a result of 
the locations of these stars on the spectroscopic mask (close to one of 
the edges).  Nevertheless, H$\epsilon$ is cleanly measured is both stars 
and so we measure $T_{\rm eff}$ and log~$g$, although these parameters will 
have larger errors.  For WD~10, the best fit model does not agree with the 
shape of the H$\beta$ line which may be contaminated.  We discuss this object 
further below.  Finally, WD~13 shows five Balmer lines (H$\beta$ -- H$8$) 
even though this star is our faintest white dwarf and therefore the spectrum is 
somewhat noisier.  Again, the measurements for this star will have larger 
uncertainties than the other higher signal-to-noise data.


\begin{table*}
\begin{center}
\caption{}
\begin{tabular}{lccccrcccccc}
\hline
\hline
\multicolumn{1}{c}{ID} & \multicolumn{1}{c}{$\alpha_{J2000}$} & 
\multicolumn{1}{c}{$\delta_{J2000}$} & \multicolumn{1}{c}{$V$} & 
\multicolumn{1}{c}{$\sigma_V$} & \multicolumn{1}{c}{$B-V$} & 
\multicolumn{1}{c}{$\sigma_{B-V}$} & \multicolumn{1}{c}{$T_{\rm eff}$ (K)} & 
\multicolumn{1}{c}{log~$g$} & \multicolumn{1}{c}{$M$ ($M_\odot$)} & 
\multicolumn{1}{c}{$t_{\rm cool}$$^{a}$ (Gyr)}\\
\hline
WD~1        & 19:20:48.6 &  37:45:48.4 & 21.91 & 0.03 & $-$0.12 & 0.04 & 34,700 $\pm$ 100 & 7.30 $\pm$ 0.03 & 0.39 $\pm$ 0.02 & $<$0.14 \\ 
WD~2$^{b}$  & 19:21:04.1 &  37:44:43.3 & 22.19 & 0.03 &  0.00 & 0.04 & 19,400 $\pm$ 100 & 7.88 $\pm$ 0.02 & 0.54 $\pm$ 0.01 & 0.063 $\pm$ 0.001 \\ 
WD~3$^{b}$   & 19:21:10.5 &  37:45:51.2 & 22.68 & 0.04 &  0.28 & 0.06 & 13,000 $\pm$ 400 & 7.80 $\pm$ 0.04 & 0.49 $\pm$ 0.02 & 1.01 $\pm$ 0.82 (0.25  $\pm$ 0.02)  \\ 
WD~4        & 19:20:58.4 &  37:45:55.5 & 23.47 & 0.08 &  0.09 & 0.11 & 17,100 $\pm$ 200 & 7.50 $\pm$ 0.04 & 0.40 $\pm$ 0.02 & 0.48 $\pm$ 0.39   \\ 
WD~5$^{b}$   & 19:20:47.3 &  37:44:37.3 & 22.82 & 0.03 &  0.15 & 0.05 & 12,500 $\pm$ 300 & 7.76 $\pm$ 0.08 & 0.47 $\pm$ 0.03 & 1.23 $\pm$ 1.01 (0.26  $\pm$ 0.03)  \\ 
WD~6        & 19:20:48.2 &  37:47:18.1 & 23.28 & 0.06 &  0.35 & 0.10 & 21,500 $\pm$ 500 & 7.33 $\pm$ 0.07 & 0.36 $\pm$ 0.02 & $<$0.53	     \\ 
WD~7        & 19:20:42.5 &  37:44:12.9 & 23.54 & 0.07 &  0.34 & 0.12 & 14,800 $\pm$ 300 & 7.91 $\pm$ 0.06 & 0.53 $\pm$ 0.02 & 0.15 $\pm$ 0.02   \\ 
WD~8        & 19:21:13.6 &  37:43:20.0 & 23.59 & 0.08 &  0.22 & 0.12 & 18,200 $\pm$ 300 & 7.73 $\pm$ 0.06 & 0.48 $\pm$ 0.02 & 0.40 $\pm$ 0.31 (0.07 $\pm$ 0.01) \\
WD~9        & 19:20:56.9 &  37:44:15.2 & 23.60 & 0.09 &  0.33 & 0.14 & 16,100 $\pm$ 300 & 7.71 $\pm$ 0.06 & 0.47 $\pm$ 0.03 & 0.57 $\pm$ 0.38 (0.11  $\pm$ 0.01)  \\ 
WD~10       & 19:20:47.0 &  37:46:29.0 & 23.80 & 0.09 &  0.03 & 0.14 & 27,700 $\pm$ 600 & 7.52 $\pm$ 0.11 & 0.44 $\pm$ 0.04 & $<$0.23           \\  
WD~11       & 19:21:05.8 &  37:46:51.5 & 23.87 & 0.16 & $-$0.14 & 0.20 & ------------     & ------------    &  ------------   & ------------      \\ 
WD~12       & 19:21:02.9 &  37:47:27.0 & 23.97 & 0.11 &  0.24 & 0.15 & 17,600 $\pm$ 600 & 7.50 $\pm$ 0.13 & 0.40 $\pm$ 0.05 & 0.48  $\pm$ 0.44  \\ 
WD~13       & 19:21:08.3 &  37:44:30.2 & 24.02 & 0.12 &  0.32 & 0.19 & 14,000 $\pm$ 500 & 7.40 $\pm$ 0.10 & 0.36 $\pm$ 0.02 & 1.09  $\pm$ 0.94  \\
WD~14       & 19:21:06.5 &  37:44:10.5 & 24.04 & 0.12 &  0.24 & 0.18 & ------------     & ------------    &  -------------- & ------------      \\
\hline
\end{tabular}
\tablenotetext{$^a$}{Cooling ages calculated using helium core models, except for WD~2 and WD~7.  Ages with carbon-oxygen core models for stars with $M \geq$ 0.47~$M_\odot$ in brackets.}
\tablenotetext{$^b$}{Possible non-cluster white dwarfs.}
\label{table2}
\end{center}
\end{table*}


The derivation of masses of white dwarfs from modeling the hydrogen 
Balmer lines has been shown to yield consistent results when compared 
to independent mass measurements, such as from gravitational redshifts 
\citep{bergeron95}.  We determine the mass for each white dwarf by 
interpolating the $T_{\rm eff}$ and log~$g$ within the updated 
evolutionary models of \cite{fontaine01}. Our standard model has a 
surface hydrogen layer mass fraction of $q(\rm H)$ = $M_{\rm H}/M$ 
= 10$^{-4}$ and helium layer of $q(\rm He)$ = 10$^{-2}$.   For the 
uncertainties in the masses, we note that if these white dwarfs are 
the products of strong mass loss on the red giant branch, they may be 
less massive than typical field white dwarfs. Surface gravities of 
less massive white dwarfs can be sensitive to the adopted hydrogen
layer thickness, and so we have calculated a new suite of low-mass,
helium core white dwarf models, using the models of \cite{hansen98}, and 
considering a full range of $q(\rm H)$ up to very thick layers, 
$q(\rm H)$ = 10$^{-2}$. Therefore, we determine the {\it range} 
of acceptable masses by considering this full range of $q(\rm H)$ in 
addition to the errors on $T_{\rm eff}$ and log~$g$.

We find that the mean mass of the twelve white dwarfs in our sample is 
0.44~$M_\odot$.  Three of the stars have masses below 0.40 $M_\odot$, five 
of the stars have masses of 0.40 -- 0.47~$M_\odot$, and only four objects 
have masses greater than 0.47~$M_\odot$.  The uncertainties on the 
individual mass measurements are typically 0.02~$M_\odot$ and 
at worst 0.05 $M_\odot$ for one star.  These results clearly 
suggest that the white dwarf population of NGC~6791 is indeed notably 
undermassive when compared to both other star clusters 
and the field distribution (see below).  As we discussed earlier, this 
is likely linked to the evolution of the progenitors of these 
white dwarfs.

We summarize the derived parameters for each white dwarf in Table~2.  We 
noted above that the best fit model for WD~10 did not reproduce the H$\beta$ 
line well.  As the mass for that star is 0.44~$M_\odot$, ignoring it from 
the sample would not change the results.  Also included in Table~2 is the 
cooling age of each star (last column).  The default values are those derived 
using the models described above 
for helium cores, except for WD~2 and WD~7.  These two stars both have 
$M$ $>$ 0.50~$M_\odot$ and therefore we have used the standard 50/50 
carbon-oxygen core models from \cite{fontaine01} to derive ages.  For four 
other white dwarfs with $M$ $\lesssim$ 0.50~$M_\odot$, in addition 
to the ages derived from helium core models we have also 
indicated the ages assuming the carbon-oxygen models in brackets.  The 
uncertainties on the cooling ages, especially for the low mass stars, 
are large as we have considered a full range in the mass of the H 
layer as discussed above.

\subsection{Confirming Cluster Membership}\label{membership}

We noted earlier in \S\,\ref{CMD} that a blank field of equal 
area taken from the outer CCDs shows a very low density of faint-blue 
stellar objects.  The expected contamination from such field white dwarfs 
in our CMD is approximately three objects.  This is $\sim$20\% of the 
number of stars targeted in our spectroscopic observations.  The masses 
of the white dwarfs derived above support this.  They are much lower 
than typical field white dwarfs and therefore these stars must belong 
to the cluster.  For example, the mass distribution of the white dwarf 
sample in the Palomar Green (PG) Survey \citep{liebert05} peaks at a mass near 
0.6~$M_\odot$.  For comparison to our NGC~6791 cluster white dwarfs, this 
sample of nearly 350 white dwarfs contains less than 25\% stars with 
$M <$ 0.54~$M_\odot$, 10\% with $M <$ 0.47~$M_\odot$, and 2.6\% with $M <$ 
0.40~$M_\odot$.  An independent estimate can be drawn from the much larger 
Sloan Digital Sky Survey, which now contains over 7000 white dwarfs in 
total \citep{kepler07}.  For those 2896 stars with $g'$ $<$ 19 (the spectral 
quality of white dwarfs in this sample is poorer than in the PG sample), 
the Sloan dataset contains 16\% stars with $M <$ 0.54~$M_\odot$, 6.3\% 
with $M <$ 0.47~$M_\odot$, and 3.3\% with $M <$ 0.40~$M_\odot$.

We can attempt to quantify which of our white dwarfs are field stars, 
if any.  For this, we first calculate a theoretical color for 
each white dwarf using the \cite{fontaine01} models and our measured 
values of $T_{\rm eff}$ and log~$g$.  Comparing this color directly to 
our $B-V$ photometry yields an estimate for the reddening of 
each star.  This reddening, coupled with an estimate of the 
star's absolute magnitude (similarly calculated from the models), 
yields the estimated distance modulus for each star.  For almost 
every white dwarf, the error in this distance modulus is 
dominated by the uncertainty in the extinction given the typical 
$\gtrsim$0.1 color error.  Cluster membership can now be established 
by comparing these distance moduli and reddenings, for each star, to 
estimates for NGC 6791. 

We find that nine of our twelve white dwarfs are consistent within the 
2-$\sigma$ range of cluster parameters.  This suggests a 25\% contamination 
fraction, slightly larger than our estimate based on the blank field earlier 
in \S\,\ref{CMD}.  Furthermore, all three objects that do not agree 
with the range of NGC~6791's distance moduli and reddening are at the high 
mass end of our sample, WD~2 (0.54~$M_\odot$), WD~3 (0.49~$M_\odot$), and 
WD~5 (0.47~$M_\odot$).  This latter object was also shown earlier to perhaps 
be in a binary system.  Therefore, the mean mass of our 
sample of white dwarfs decreases to 0.43~$M_\odot$ if we exclude these 
three possible field white dwarfs.  However, we note that two of the three 
excluded stars have a mass significantly less than the field distribution and 
therefore it is not definitive that they are non-members.  The method 
used to estimate membership here is approximate and does not take into 
account all possible biases.  For example, small uncertainties in 
the theoretical colors and magnitudes from the white dwarf models are 
ignored and there may even be increased intrinsic extinction around 
these white dwarfs due to the progenitor mass loss.


\section{Discussion} \label{discussion}

\subsection{The Extreme Horizontal Branch of NGC~6791} \label{ehbstars}

\begin{figure*}
\begin{center}
\leavevmode 
\includegraphics[height=17cm,angle=270]{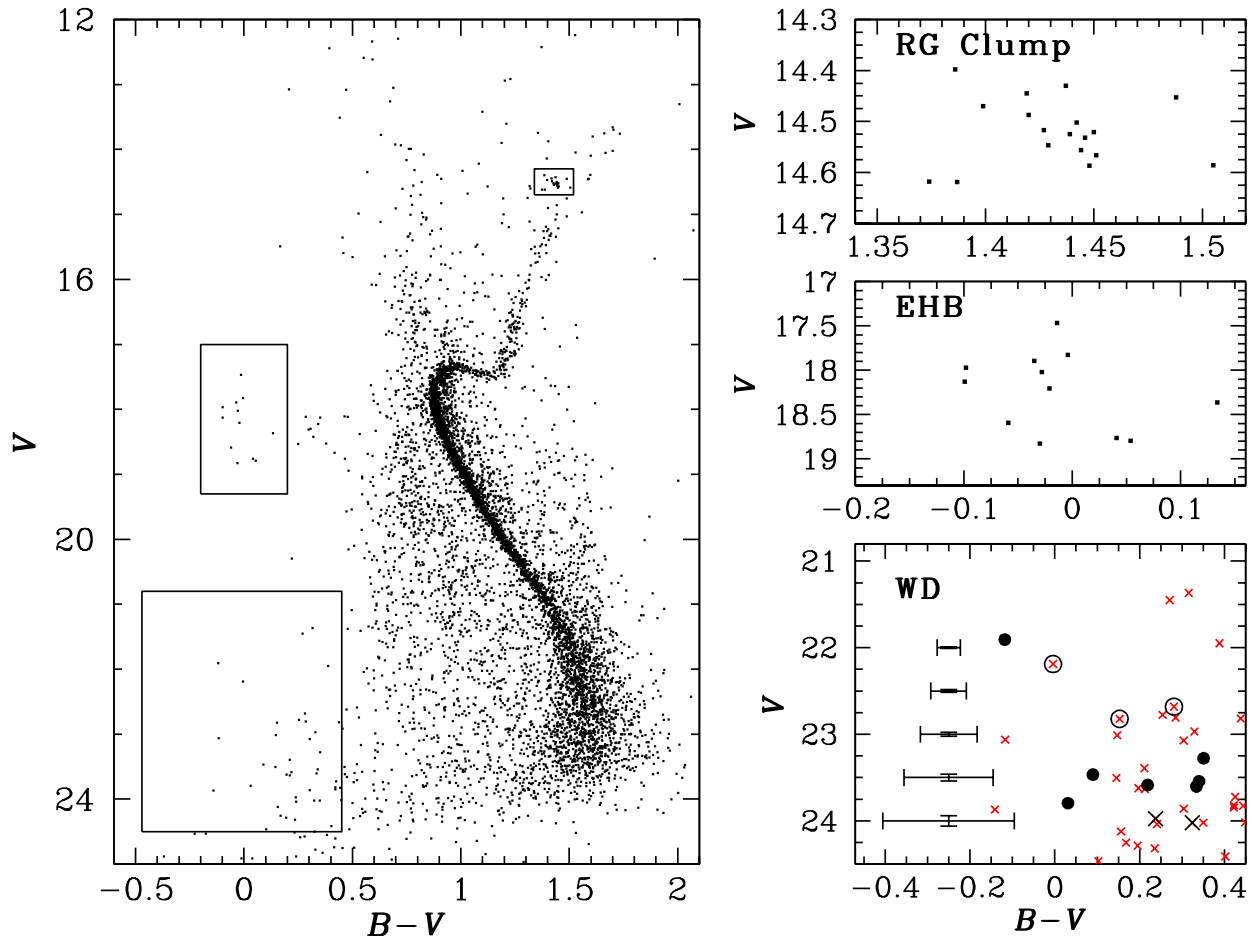}
\end{center}
\caption{A closer look at the red giant clump (RG clump -- 
top-right) and the extreme horizontal branch (EHB -- middle-right) 
of NGC~6791 reveals approximately a dozen stars in each 
phase.  The white dwarf cooling sequence is also shown in the 
bottom-right panel, along with an indication of the photometric 
errors in the data.  The larger filled (open) circles mark the locations 
of the confirmed cluster (possible field) white dwarfs in this study.  To 
help illustrate the locations of these three post main-sequence evolutionary 
phases on the full CMD, we mark boxes on the left-panel corresponding to 
these zoomed regions.
\label{fig:cmdzoom}}
\end{figure*}

The CMD of NGC~6791 (Figure~\ref{fig:cmd}) clearly shows both a red giant 
clump and an extremely blue horizontal branch as discussed earlier. In 
Figure~\ref{fig:cmdzoom} we take a closer 
look at these two phases, as well as the white dwarf cooling sequence 
of the cluster.  In the top-right and middle-right panels, we count 
a total of approximately a dozen stars that are in each of the red giant clump 
and extreme horizontal branch phases of evolution (over our field area).  The 
presence of both of these core helium burning phases likely suggests 
that the red giants have undergone stochastic mass loss.  In fact, the 
extremely blue horizontal branch is a likely sign that a fraction of the stars 
in this cluster have lost an increased amount of mass relative to the 
``normal'' evolution that creates the red giant clump.


An alternate method of producing extreme horizontal branch stars involves 
binary evolution in which one star loses mass to a companion (see e.g., Han 
et~al.\ 2003).  However, searches for binary companions among globular cluster 
extreme horizontal branch stars have been unsuccessful \citep{monibidin06a,monibidin06b}.  
Such a scenario is also not likely in NGC~6791.  \cite{janes97} examined the CMDs of 
about a dozen mostly old open clusters and found that NGC~6791 contains the lowest 
binary fraction of the group, 14\%.  The mean fraction among the rest of the 
sample is 30\%.  Qualitatively, a large binary fraction for NGC~6791 
appears to be ruled out from our much deeper CMD as well.  There is no evidence 
for an obvious equal mass binary sequence nor a very strong signature of extra 
scatter above the cluster main sequence relative to the CMDs of other rich clusters 
such as NGC~6819, NGC~2099, NGC~2168, and NGC~2323.  All of 
these other clusters have been shown to contain 20 -- 30\% binaries through 
synthetic CMD tests \citep{kalirai04}.  If binary evolution is the cause of the 
extreme horizontal branch, then it is very unusual that these other clusters do 
not contain any stars in this phase.  In fact, the only other open cluster that 
shows evidence for an extreme horizontal branch happens to be very similar to 
NGC~6791 in its fundamental properties.  NGC~188 is both an old and metal-rich 
system and contains two of these hot stars \citep{dinescu96}.  This strengthens 
the case for a metallicity-related origin of the extreme horizontal branch stars 
in these clusters.  Binarity also suggests that the extreme 
horizontal branch stars in NGC~6791 should be centrally concentrated and 
should contain a significant spread in luminosity, neither of which are observed 
\citep{liebert94}. The derived luminosity range is in fact consistent with that 
expected from metal-rich, hot horizontal branch stars \citep{landsman98}.

Direct photometric and spectroscopic probes to confirm the nature of the extreme horizontal 
branch stars in NGC~6791 and NGC~188 have largely been unsuccessful.  \cite{chaboyer02} 
obtained far ultraviolet images with the Space Telescope Imaging Spectrograph on 
{\it HST} to study the possible progenitors of the extreme horizontal branch stars, the
bluest of the giant branch stars.  If the binary formation theory is correct, then a large 
fraction of these giants should contain white dwarf companions which could potentially be 
seen in the ultraviolet.  However, in a dozen targeted stars (six in each cluster), 
none of the NGC~6791 giants and just two of the NGC~188 giants showed a far ultraviolet 
flux (which may itself come from the chromosphere of the giant star).  Detailed 
abundance analysis of the coolest extreme horizontal branch star in NGC~6791 
combined with its optical colors favors it having suffered from heavy line blanketing 
due to the high metallicity as opposed to a binary nature.  Although \cite{green97} do 
find that two of the other NGC~6791 horizontal branch stars are spectroscopic binaries, 
these two systems are not extremely blue horizontal branch stars.

Taken together, this evidence suggests that the likely cause of the extreme horizontal 
branch in NGC~6791 is related to the high metallicity of the cluster and not binary 
evolution.  High dispersion observations of the fainter extreme horizontal branch stars 
(as obtained for the blue horizontal branch stars) could provide the definitive answer.  

\subsection{Avoiding Core Helium Burning} \label{avoiding}

The spectroscopic mass measurement of NGC~6791's white dwarf population 
indicates that in addition to a red giant clump and extreme horizontal branch, 
there is yet a third, even more radical, evolutionary channel for the stars of 
this cluster.  Table~2 indicates that two-thirds of the NGC~6791 {\it member} white 
dwarfs have masses below the threshold ($\sim$0.46~$M_\odot$) at which 
helium is ignited to produce a carbon-oxygen mixture in the core.  This suggests 
that the progenitor red giants of these stars did not experience 
a helium flash and therefore bypassed both of the above phases and landed 
directly on the white dwarf cooling sequence (with helium cores).  Such evolution 
is consistent with models of red giants that suffered extreme mass loss (see 
section~\ref{theoryRGB} -- D'Cruz et~al.\ 1996).  It is also worth noting 
that all 12 of the NGC 6791 white dwarfs are DA spectral type.  Based on 
the field white dwarf ratio, we would statistically expect a few of these 
stars to be DB (helium atmosphere) white dwarfs.  A possible explanation for 
this may be related to the unique evolutionary paths of the progenitor stars 
which avoided the shell helium burning phase.

The cumulative effect from the post main-sequence evolution of {\it all} stars 
in NGC~6791 is shown in the bottom-right panel of Figure~\ref{fig:cmdzoom}.  The 
crosses mark all objects on the CMD and the filled (open) circles mark 
the confirmed (possible field) cluster white dwarfs.  Not surprisingly, the bright 
part of the white dwarf cooling sequence looks unlike that of other star 
clusters, showing much more scatter.  For example, the sequences of the open 
clusters M67 \citep{richer98} and NGC~6819 \citep{kalirai01b}, as well as 
the globular clusters M4 \citep{hansen04} and NGC~6397 \citep{richer06} 
exhibit a tighter distribution of points in the faint-blue end of the 
CMD.  Several factors likely contribute to the scatter.  First, we noted 
in \S\S\,\ref{CMD} \& \ref{membership} that up to three of the white 
dwarfs in our sample could potentially be field white dwarfs and therefore 
there may be a 20 -- 30\% contamination fraction among all objects (crosses).  
Second, the masses of the {\it cluster} white dwarfs, and therefore their 
core compositions, are different along the cooling sequence.  The evolutionary 
timescales of these stars therefore vary and this would work to wash out a 
tight cooling sequence.  However, if this were the only cause we should see 
a correlation between the white dwarf masses and their positions in the CMD.  
Figure~\ref{fig:wdspectra} shows that this is, in general, not the case.  
Although it can not be a large effect for the reasons outlined above, some 
binary evolution may be present in our white dwarf sample.  The spectra of 
both WD~5 and WD~10 show evidence of contamination, possibly 
from faint companions.  Any mass transfer in the evolution of these systems 
would certainly alter the subsequent evolution on the white dwarf cooling 
sequence (see e.g., Hurley \& Shara 2003).  Although statistically unlikely, 
it is also possible that we have targeted a double degenerate system.  Finally, 
we have plotted both a $V$ and $B-V$ photometric error bar at different 
magnitudes in the bottom-right panel of Figure~\ref{fig:cmdzoom}.  The results 
show that for $V >$ 23 the errors in our colors are comparable to the spread 
seen in the CMD.  This suggests that our photometric errors are also likely 
dominating the scatter observed on the CMD.

A much better test of the true intrinsic spread along the NGC~6791 white 
dwarf cooling sequence can be judged from the deep {\it HST}/ACS CMD of 
this cluster \citep{bedin05}.  These data are not affected by photometric 
errors at these magnitudes.  The \cite{bedin05} CMD shows clear evidence for 
a scatter of 0.25 -- 0.30 magnitudes (in color) near the tip of the cooling 
sequence and extending all the way down to the faintest white dwarfs.  
This rules out photometric errors and therefore the observed spread must 
be related to the various evolutionary channels that have led to the 
formation of these stars, the root of which is the mass loss on 
the red giant branch.  Interestingly, \cite{bedin05} find that the location 
of the reddest white dwarfs along their cooling sequence is consistent 
with pure helium core models of low mass (0.32 $M_\odot$).  As we saw 
in \S\,\ref{WDMasses}, the observed spread in masses of the NGC~6791 
white dwarfs ranges from 0.36 -- 0.54~$M_\odot$ and therefore the 
dominant bluer sequence of white dwarfs in their CMD (that they fit with 
carbon-oxygen core models to derive the young age) actually contains 
a mixture of these canonical white dwarfs (those with progenitors in the 
red giant clump) and more massive helium core white dwarfs.  In our 
sample of bright white dwarfs, WD~7 ($M$ = 0.53~$M_\odot$) likely 
represents a star that evolved through this normal channel.

\subsection{Red Giant Branch Mass Loss -- \\ Theoretical Estimates} \label{theoryRGB}

The evolutionary channel discussed above requires some fraction of the stars in 
NGC~6791 to have experienced enhanced mass loss during their evolution.  There 
are three primary mechanisms for the total post-main sequence mass 
loss in stars: stationary winds, dust related outflows, and pulsation related outflows 
(e.g., Willson 2000).  The majority of the mass loss takes place while a star 
is ascending the asymptotic giant branch and evolving through the planetary 
nebula phase, although the star will also lose an appreciable amount of mass 
on the red giant branch.  It is not well understood whether this latter mass 
loss, i.e., that occurs prior to the horizontal branch phase, is driven 
primarily via winds on the red giant branch itself or as a result of the 
helium flash.  However, the amount of the red giant branch mass loss, is a 
sensitive function of the stellar metallicity, as chemically enriched stars will 
lose a larger fraction of their total mass.  

To estimate the expected mass loss along the red giant branch, we invoke the 
models of \cite{marigo01}.  These models provide chemical yields for both 
low- and intermediate- mass stars evolving from the zero age main sequence to 
the end of the thermally pulsating asymptotic giant branch.  The integrated mass 
loss for a slightly metal-poor ([Fe/H] = $-$0.7), 1.05 $M_\odot$ star 
(appropriate mass for an NGC~6791 giant) is 41\% of its initial mass.  
A Solar metallicity star of the same mass will lose 48\% of its mass through its 
evolution.  However, $\sim$40\% of the Solar metallicity star's mass loss 
will occur on the red giant branch whereas 33\% of the [Fe/H] = $-$0.7 star's 
mass loss occurs on the red giant branch.  For a metallicity as extreme as 
NGC~6791's ([Fe/H] = $+$0.3 -- $+$0.5), a star will lose even a larger fraction of 
its mass on the red giant branch. \cite{cruz96} estimate that a 1.08 $M_\odot$ 
star with [Fe/H] = $+$0.37 will form a core with a mass of just 0.45 -- 
0.47~$M_\odot$.

These theoretical calculations suggest that the amount of mass loss along 
the red giant branch of NGC~6791 will yield a final mass of the star at 
the tip of the branch that is within a few-hundredths of the critical mass 
needed to ignite helium in the core.  Given the stochastic nature of the 
red giant branch mass loss, some stars in NGC~6791 certainly reached the 
critical mass whereas others did not.  The large internal 
metallicity dispersion within the cluster (rms = 0.08 dex -- Gratton 
et~al.\ 2006) will also add to the variable mass loss.  For example, 
\cite{worthey03} present low-resolution spectra of K giants and find 
that one star in this cluster has an extremely high metal abundance, 
[Fe/H] = $+$0.6.  The theoretical arguments for this mass loss are 
therefore qualitatively consistent with our conclusions above based 
on the morphology of the NGC~6791 CMD and the masses of the cluster 
white dwarfs.  

\subsection{The Luminosity Function of NGC~6791's \\ Red Giant Branch} \label{obsRGB}

If, in fact, a significant fraction of NGC~6791's stellar population is peeling away 
from the red giant branch before the helium flash, then the luminosity function 
of the cluster's red giant branch should be depleted as one approaches the tip 
(see e.g., Sandquist \& Martel 2007).  An analysis of the cluster's red giant 
branch by \cite{garnavich94} found that its tip does not rise above 
$M_I \sim$ $-$2.7, over a magnitude fainter than metal-rich globular 
clusters.  Interestingly, the recent study of \cite{luck07} compares the 
metallicity distribution functions of nearby field dwarfs and giants, and 
finds that the giant distribution lacks a metal-rich tail.  To test whether 
there is a {\it thinning} out of this upper red 
giant branch, we compare the cluster's red giant branch luminosity function 
to that of three other old open star clusters, Berkeley~17 (8.5 Gyr -- 
Bragaglia et~al.\ 2006), M67 (4.3 Gyr -- Richer et~al.\ 1998), and NGC~188 (6.8 
Gyr -- Stetson, McClure, \& VandenBerg 2004 and references within).  We isolate 
the red giant branch stars from the published CMDs in these studies and apply 
the derived distance moduli to each data set.  We also confirmed that our study 
is not incomplete near the tip of the red giant branch, where these stars become 
increasingly redder.  For this, we matched our optical data to the near infrared 
study of \cite{carney05} and were able to recover all of the red giants near 
the tip.


\begin{figure}
\epsscale{1.1} \plotone{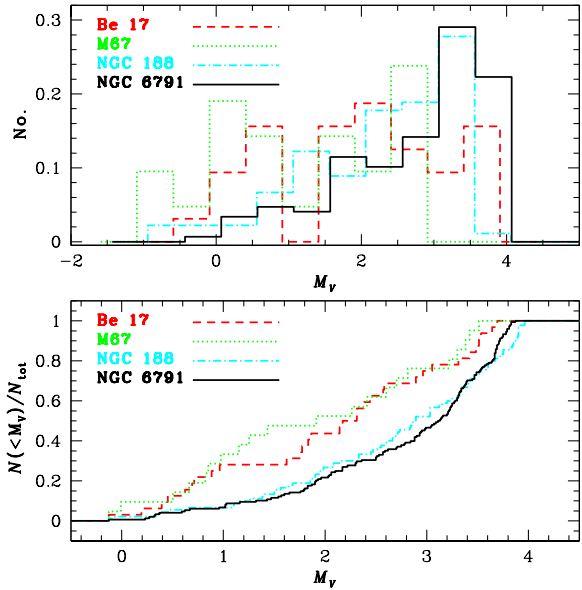} 
\figcaption{The differential (top) and cumulative (bottom) red giant 
branch luminosity function of NGC~6397 (solid) is compared to those of 
three other old open clusters, Berkeley~17 (dashed), M67 (dotted), 
and NGC~188 (short dash-dot).  Both panels indicate that the number of 
red giants in NGC~6791 decreases more rapidly than the other clusters 
as the tip is approached.  As discussed in the text, this {\it thinning} 
out of the upper red giant branch suggests that stars are peeling away, 
never having experienced a helium flash, and forming undermassive 
helium core white dwarfs.
\label{fig:LF}}
\end{figure}


Figure~\ref{fig:LF} (top) shows the red giant branch luminosity function for 
each cluster.  We have plotted this with the tip of the branch towards 
the left of the diagram.  Both in NGC~188 (also a metal-rich cluster) and NGC~6791, 
the luminosity functions are heavily skewed towards the base of the 
red giant branch.  The decline in the number of stars as the tip is approached 
is more rapid in NGC~6791 than in all three other clusters.  To illustrate 
this more clearly, we build cumulative luminosity functions for each cluster in 
Figure~\ref{fig:LF} (bottom) and adjust each function so the brightest observed 
giant is shifted to a common magnitude.  This aligning of the tip of the red 
giant branch has the advantage that it is largely independent of both chemical 
composition and age \citep{sandquist07}.  The NGC~6791 luminosity function is clearly 
``bottom-heavy'' relative to the other clusters, suggesting that there is 
an absence of red giants near the tip.  This absence is not caused by very red 
stars that may have escaped detection in our study. To test this further, we apply a 
Kolmogorov-Smirnov (K-S) test \citep{press86} between the NGC~6791 luminosity 
function and each of the other three clusters.  The K-S test makes no assumption 
about the distribution of the data and is therefore insensitive to possible 
spurious biases from arbitrary binning of data.  The significance level 
probability, $P$, that the null hypothesis is true (i.e., that the luminosity 
functions are drawn from the same distribution) is found to be $<$0.01 when 
comparing NGC~6791 to each of Berkeley~17 and M67.  For the two more metal-rich 
clusters with extreme horizontal branches, NGC~6791 and NGC~188, the K-S test yields 
a probability $P$ = 0.31 (i.e., it can not be shown that these two luminosity 
functions are drawn from different distributions).  


\subsection{Resolving the Age Discrepancy of NGC 6791}\label{agediscrepancy}

The observed white dwarfs in our sample are the brightest objects along 
the cooling sequence whereas most of the leverage in measuring the white dwarf 
cooling age of the cluster comes from fainter white dwarfs in the \cite{bedin05} peak.  
Over these magnitudes, stars along this cooling sequence have a very small change 
in mass (also true for their progenitors).  Therefore, the same fate should affect 
all of these stars unless we have not accounted for a piece of missing physics.  
\cite{hansen05} estimated the number of white dwarfs in the 
Bedin et~al.\ CMD that came from each of the three evolutionary channels by 
considering the lifetimes of stars in the red giant clump, extreme horizontal 
branch, and white dwarf cooling phases, and by assuming a ratio of extreme 
horizontal branch stars to red giant clump stars in those data.  
The CMD presented in Figure~\ref{fig:cmdzoom} overlaps the 
{\it HST}/ACS study and so we can now refine this calculation.  Over the ACS 
field targeted by Bedin et~al., the CMD of NGC~6791 shows two extreme horizontal 
branch stars and four red giant clump stars.  The evolutionary timescale for these 
core helium burning phases lasts approximately 1 $\times$ 10$^{8}$ years, whereas 
the cooling time for the faintest white dwarfs in the HST study is $\sim$2 $\times$ 
10$^{9}$ years (for carbon-oxygen core models down to $F606W$ = 28).  Therefore, scaling 
arguments suggest that 120 of the white dwarfs in the Bedin et~al.\ study should 
have evolved from either of these two evolutionary channels and therefore 
have carbon-oxygen cores (two-thirds of which came from the canonical red giant clump 
phase).  The white dwarf cooling sequence of the cluster shows roughly 600 
white dwarfs, and therefore up to 80\% of the white dwarfs may be helium core 
stars.

\begin{figure}
\epsscale{1.1} \plotone{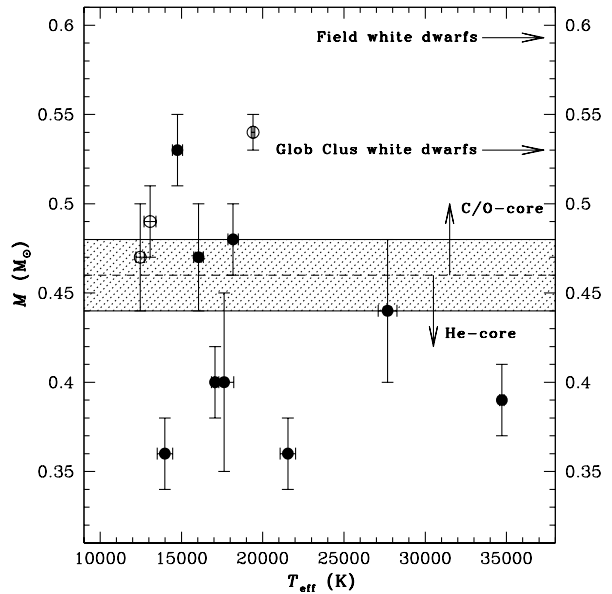} 
\figcaption{The $M$--$T_{\rm eff}$ plane for all confirmed 
NGC~6791 white dwarfs (solid points) as well as those that 
were determined to be 2-$\sigma$ outliers from cluster membership 
in \S\,\ref{membership} (open circles).  Six of the nine cluster 
white dwarfs clearly have a mass under the threshold (hashed region -- 
$M \sim$ 0.46~$M_\odot$) at which helium would have been ignited 
in the core of the progenitor star to produce carbon-oxygen.  The 
arrow marks the mean mass of the field white dwarf distribution 
from the Sloan Digital Sky Survey (0.59 $M_\odot$) and the mean 
mass of the globular cluster NGC~6752's white dwarf population 
(0.53 $M_\odot$ -- Moehler et~al.\ 2004).  All of our cluster 
white dwarfs have a mass less than these values.
\label{fig:massteff}}
\end{figure}

An independent check on these numbers can now be derived from our mass 
measurements.  Figure~\ref{fig:massteff} indicates that the ratio of 
helium-core to carbon-oxygen core white dwarfs in NGC~6791 is $\sim$2:1, 
assuming that 0.46~$M_\odot$ is the threshold at which a carbon-oxygen core 
will be produced in the progenitor star.  If this critical mass is slightly 
larger, i.e., 0.48~$M_\odot$, then the ratio may be as large as 8:1.  Since 
the cooling is a factor of $\sim$3$\times$ slower for the helium core stars, 
their cluster birthrate is then estimated to be between 40 -- 70\% of all 
stars.  


The fraction of helium core vs carbon-oxygen core 
white dwarfs directly affects the inferred cluster white dwarf cooling 
age as the cooling rates are proportional to the mass number 
(chemical species) of the core.  For example, for a fixed stellar mass, 
a 50/50 carbon-oxygen mixture will have $\sim$3.5$\times$ fewer ions than 
a helium core star.  This results in a higher heat capacity for the helium 
core white dwarf and therefore the cooling will be $\sim$3.5$\times$ slower 
than that of a carbon-oxygen core white dwarf of similar mass.  This is of 
course offset slightly by the fact that the mean mass of the helium core 
stars is somewhat smaller than that of normal carbon-oxygen core white dwarfs 
(a relatively small effect in this case).  By comparing the observed white dwarf 
luminosity function to model luminosity functions that invoke pure 
helium cores, pure carbon-oxygen cores, and a combination of the two, 
\cite{hansen05} was able to show that an age of 7--8~Gyr requires the 
birthrate fraction of helium core white dwarfs in the cluster to be 
between 50\% and 100\% of the total white dwarf production.  This is 
consistent with our findings above.


\subsection{The Ultraviolet Upturn in Elliptical Galaxies}\label{uvupturn}

The ultraviolet upturn in elliptical galaxies and spiral bulges refers 
to the sharp, rising flux in the integrated spectra of 
these objects shortward of $\lambda \sim$ 2500 ${\rm \AA}$ 
\citep{burstein88,greggio90,bressan94,oconnell99,brown04}.  The 
phenomenon is found to be extremely variable in different 
galaxies, and therefore must be very sensitive to the properties 
of the sources responsible for the upturn.  The dominant, old 
main-sequence stellar population of these galaxies is too faint, and cool, to 
constitute any appreciable fraction of this flux.  Therefore, the sources must either 
be young massive stars (e.g., in active star formation regions) or more 
exotic populations of low-mass, hot stars.  Recent results from GALEX suggest 
that only in rare cases does the former dominate the ultraviolet flux 
\citep{salim05,rich05,donas06}.  For the latter, \cite{greggio90} 
consider the energetics of several possible sources including binary candidates, 
hot or accreting white dwarfs, post asymptotic giant branch stars, and extreme 
horizontal branch stars and find that the extreme horizontal branch stars are 
the most likely candidates (although, see recent study by Han, Podsiadlowski, 
\& Lynas-Gray 2007 suggesting a binary model for the excess).  These stars are hot, 
bright, and can establish an equilibrium while burning helium in their cores 
for $\sim$10$^8$ years.  Unfortunately, resolved photometry, or spectroscopy, 
of the sources for the ultraviolet upturn in these distant galaxies is currently 
not possible (although M31 and M32 have been imaged in the ultraviolet - Bertola 
et~al.\ 1995; Brown et~al.\ 2000).   

In Galactic globular clusters extremely blue horizontal branches have long 
been argued to exist because of the system's old age, low metallicity, and 
differential mass loss on the red giant branch \citep{iben70}.  However, 
the properties of these systems (e.g., low metallicity, high density) are 
contrary to the extragalactic populations \citep{liebert94}.  Recent studies 
have shown that metal-rich disk clusters also possess an extended blue horizontal 
branch (e.g., Rich et~al.\ 1997), thereby providing nearby systems, with some 
similar properties to the elliptical galaxies and spiral bulges, that possibly 
harbor the sources causing the ultraviolet upturn.  Given its old age, very high 
metal abundance, and small distance modulus, NGC~6791 represents a nearby 
analog of the metal-rich component in these galaxies that can be used to 
understand the origin of the extra far-ultraviolet radiation (see also 
Liebert, Saffer, \& Green 1994).

We have shown that the white dwarf population of NGC~6791 is undermassive, 
a consequence of enhanced mass loss in the evolution of the progenitors of 
these stars.  The extreme horizontal branch stars of NGC~6791 are a subset 
of this population, those that lost mass on the red giant branch but not 
enough to prevent the ignition of helium while the star was well on its 
way to becoming a white dwarf.  This type of evolution clearly suggests 
that even within a co-eval system with little metallicity spread, the mass loss 
is variable.  Of course this effect would be compounded in a system such as 
an elliptical galaxy which has a large spread in both metallicity and age 
relative to an open cluster.  Furthermore, as we mentioned earlier, the 
amount of ultraviolet flux in these galaxies is found to vary enormously 
from galaxy to galaxy.  The variable excess likely depends sensitively 
on the number of extreme horizontal branch stars and could even be 
susceptible to their specific location along the extreme horizontal 
branch.  The stochastic mass loss, due to the metallicity, could therefore be 
a contributor in dictating the amount of this ultraviolet flux in 
ellipticals and spiral bulges. Other sources, such as the presence of a 
minority population of helium enhanced stars \citep{piotto05}, may also 
be important.

In total, about 30\% of the NGC~6791 helium burning stars are hot (including 
the new ones found by Buson et~al.\ 2006).  Of course, this is an extreme case 
considering the very high metallicity of the cluster; slightly lower metallicity 
populations are expected to produce a smaller fraction. \cite{dorman95} construct 
synthetic population models and find that even for the bluest galaxies, a 
fraction of just 15 -- 20\% of the stars would need to become extreme horizontal 
branch stars to explain the ultraviolet colors.  Therefore the production of 
extreme horizontal branch stars in an environment such as NGC~6791 is certainly 
sufficient to explain the ultraviolet upturn in elliptical galaxies and 
spiral bulges.  If, in fact, the amount of excess ultraviolet radiation in these 
systems increases with metallicity, then this could directly correlate with 
the picture of enhanced mass loss in metal-rich systems that would produce 
a larger fraction of the sources of the radiation.  However, the jury is still 
out on whether such a trend exists.  For example, Burstein et~al.\ (1988) see 
a positive correlation between the ultraviolet excess and metallicity whereas 
other studies do not (e.g., Rich et~al.\ 2005). 


\section{Conclusions}\label{conclusion}

We have presented the first spectroscopic observations of the white dwarf 
population of the metal-rich open star cluster NGC~6791.  Two-thirds of 
the white dwarfs in this cluster are found to be undermassive, well below 
the $M \sim$ 0.46~$M_\odot$ threshold at which helium would have ignited 
in the cores of the progenitor red giant branch stars to produce carbon-oxygen.  
These helium-core white dwarfs have formed in this cluster as a result of 
enhanced mass loss on the red giant branch, due to the high metallicity 
of the cluster ([Fe/H] $\sim$ $+$0.4).  This scenario naturally provides 
an explanation for the recently reported low white 
dwarf cooling age of NGC~6791 by \cite{bedin05}; helium-core white dwarfs 
cool a factor of $\sim$3$\times$ slower than carbon-oxygen core white 
dwarfs and therefore their measurement of 2.4~Gyr would actually imply 
a cooling age of $\gtrsim$7~Gyr, consistent with the well measured main-sequence 
turnoff age.  The enhanced mass loss also explains the presence 
of the extreme horizontal branch stars in this cluster which are plausible 
analogs of the sources responsible for the ultraviolet upturn in elliptical 
galaxies and spiral bulges.  

Interestingly, a recent study by \cite{kilic07} has reported that there is a 
significant analogous field population of low mass, single white dwarfs that arises 
from old, metal-rich stars that suffered severe mass loss and avoided the helium 
flash.  Therefore, the results in this paper are likely a general consequence of 
stellar evolution in high metallicity environments

Finally, it is worth noting that these results do not rule out that the cooling rate 
of some of NGC~6791's white dwarfs has been retarded from the sedimentation 
of $^{22}$Ne \citep{deloye02}.  $^{22}$Ne is produced during helium burning and so the 
helium core white dwarfs would be unaffected, however, this could represent a 
heating mechanism for the carbon-oxygen core stars.  The rate at which $^{22}$Ne 
settles into the core depends on internal physics that are currently not well 
understood.  Further investigation of this heating source will likely clarify how 
dominant this effect is in a metal-rich environment such as NGC~6791.  Other 
explanations related to white dwarf physics, binarity, or enhanced helium 
abundances may yet emerge.  Future observations of NGC~6791 may be able to 
confirm our picture.  Ground based observations of the cluster red giants in 
the mid-infrared could potentially show signs of the enhanced mass loss 
(e.g., through a 10 micron excess).  A deeper study of the cluster white dwarfs 
should also be undertaken with {\it HST}.  Such observations should unveil 
a second peak in the white dwarf luminosity function resulting from the cooling of 
canonical carbon-oxygen core white dwarfs.  The second epoch data could also provide 
for a much cleaner study of cluster stars through proper motion selection.


\acknowledgements
We gratefully acknowledge L.~Bildsten for reading our manuscript before submission and 
suggesting several useful additions and clarifications related to the cooling of 
white dwarfs.  We wish to also acknowledge help from D.~Leong and E.~Chudwick with data 
processing.  We are grateful to R.~Peterson and B.~Holden for insightful discussions 
related to extreme horizontal branch stars and the ultraviolet upturn in elliptical 
galaxies.  We thank S.~Kepler for kindly providing us with the list of magnitudes and 
masses for the sample of white dwarfs in the Sloan Digital Sky Survey.  JSK is 
supported by NASA through Hubble Fellowship grant HF-01185.01-A, awarded by the 
Space Telescope Science Institute, which is operated by the Association of 
Universities for Research in Astronomy, Incorporated, under NASA contract 
NAS5-26555. Support for this work was also provided by grant HST-GO-10424 
from NASA/STScI.  PB is a Cottrell Scholar of Research Corporation.  The 
research of PB and HBR is supported by grants from the Natural Sciences and 
Engineering Research Council of Canada.  HBR also thanks the Canada-US Fulbright 
Program for the award of a Fulbright Fellowship.


\end{document}